\RequirePackage{fix-cm}
\documentclass[smallextended]{svjour3}       
\smartqed  
\usepackage{graphicx}
\begin{document}

\title{Kaluza-Klein Bubble With Massive Scalar Field}

\author{Darrell Jackson}


\institute{D. Jackson \at
              Applied Physics Laboratory \\
              University of Washington \\
              Seattle, WA USA \\
              Tel.: 206-909-9678\\
              \email{drj12@uw.edu}\\
              ORCID 0000-0001-8060-2439\\
}

\date{Received: date / Accepted: date}

\maketitle

\begin{abstract}
A well-known soliton (bubble) solution of five-dimensional Kaluza-Klein General Relativity is modified by imposing mass on the scalar field. 
By forcing the scalar field to be short-range, the failure of the original bubble solution to satisfy the equivalence principle is remedied, 
and the bubble acquires gravitational mass.
Most importantly, the mass is quantized, even in this classical setting, and has a value $m_P / (4 \sqrt{\alpha})$, where $m_P$ is the Planck mass,
and $\alpha$ is the fine-structure constant. This result applies for any choice of scalar-field mass, as it is an attractor for the field equations.

\end{abstract}

\section{Introduction}
\label{Intro}
Five-dimensional Kaluza-Klein (K-K) General Relativity allows finite source-free
solutions, which have been referred to as solitons. 
Among the various solutions in
the literature, the present discussion focuses on a well-known wormhole solution
 \cite{bib1,bib2,bib3}. A bubble solution
is obtained by pinching-off of the interior region and replacing it with a hole 
in spacetime in the manner of Witten's \cite{bib4} expanding K-K ``Bubble of Nothing." 
The pinch-off condition fixes the bubble radius to be proportional to the radius of the  fifth dimension,
but the solution violates the equivalence principle, as it has zero gravitational mass and finite inertial mass \cite{bib2}.
In the present work, the bubble is modified to satisfy the equivalence principle by introducing a mass term in the
scalar equation of motion. The mass term has the additional benefit of voiding K-K scalar-tensor gravity \cite{bib5},
removing a troublesome property of the original K-K Relativity. The primary payoff is that the new bubble has a gravitational mass,
equal to about three Planck masses.

It should be noted that the fifth dimension is assumed here to be compact, unlike approaches such as ``Induced-Matter Thoery" \cite{bib3}, and
the Randall-Sundrum model \cite{bib6}. Compactness was assumed by Klein, with resulting quantization of
charge. Equally important, this assumption introduces a length scale (the radius of the fifth dimension). 

\section{The Wormhole Metric}
\label{Metric}
Following Tomimatsu \cite{bib7}, the five-dimensional line element 
for the solution of interest in spherial coordinates is
\begin{equation}
\label{Eq:line_element_5D}
ds^2 = -dt^2 + g_{rr}dr^2 + r^2 d\Omega^2 + g_{55} (dx^5)^2~,
\end{equation}
where 
\begin{equation}
\label{Eq:hrr}
g_{rr} = \frac{1}{1-r_b/r}~, 
\end{equation}
and
\begin{equation}
\label{Eq:h55}
g_{55} = 1-r_b/r~. 
\end{equation}
This metric has a long-range scalar field and is a solution of the vacuum equation
\begin{equation}
\label{Eq:Rab}
R_{ab}^{(5)} = 0~, 
\end{equation}
where $R_{ab}^{(5)}$ is the five-dimensional Ricci tensor with indices including
$x^5$. The 5-D metric components are denoted $g_{ab}$, while the 4-D metric is
$g_{\mu \nu}$ with $\mu$ and $\nu$ restricted to the usual space-time coordinates. 

The equivalent 4-D solution obeys
\begin{equation}
\label{Eq:Rmunu}
R_{\mu\nu}  = T_{\mu\nu}~, 
\end{equation}
where $R_{\mu\nu}$ is the 4-D Ricci tensor, and the line element is identical to (\ref{Eq:line_element_5D})
and (\ref{Eq:hrr}) except for the exclusion of
$g_{55}$. Expresson (\ref{Eq:Rmunu}) differs from the usual Einstein field equation because the stress-energy
tensor is traceless, as will be demonstrated shortly. The 4-D stress-energy tensor is
\begin{equation}
\label{Eq:Energy_Momentum}
T_{\mu\nu}  = \Psi_{; \mu\nu} + \Psi_{,\mu}\Psi_{,\nu}~. 
\end{equation}
As with any scalar field, $ \Psi_{; \mu\nu} =  \Psi_{; \nu\mu}$, so the stress-enegy tensor is symmetric in $\mu$ and $\nu$. The units employed here are such that the factor $8 \pi G / c^4$ normally seen in Einstein's field equations has been absorbed into the definitions of the Einstein tensor and stress-energy tensor, as it does not appear naturally in five-dimensional relativity.
The scalar field $\Psi$ is proportional to the logarithm of $g_{55}$:
\begin{equation}
\label{Eq:psi}
g_{55}  = e^{2\Psi}~, 
\end{equation}
and obeys the equation
\begin{equation}
\label{Eq:psi_wave_eq}
\Psi_{;\mu}^{~~\mu}  + \Psi_{,\mu} \Psi_;^{~\mu} = 0~. 
\end{equation}
From this it follows that the stress-energy tensor is traceless,
and energy and momentum are conserved:
\begin{equation}
\label{Eq:En_Mom_Consv}
T_{\mu\nu;}^{~~~\nu}  =0~.
\end{equation}
The coordinates $t$ and $x^5$ are cyclic, and the fifth coordinate is compact,
suggesting the introduction of an alternative fifth coordinate, $\phi$, through
\begin{equation}
\label{Eq:phi}
x^5 = a \phi~,
\end{equation}
where $a$ is the radius at infinity of the fifth dimension, and $0 \leq \phi < 2\pi$.
In isotropic coordinates \cite{bib3}, the coordinate transformation $r' = r_b^2/r$ leaves the metric invariant, but exchanges the
interior ($0 < r < r_b$) and exterior ($r_b < r < \infty$) regions, 
showing that this is a wormhole connecting two identical, asymptotically flat
spaces. The proper radius measuring the circumference and area of the wormhole is $r_b$.

Constancy of the temporal metric component shows that this solution has zero gravitational
mass, and lack of ($x^5, t$) off-diagonal terms indicates that the vector (EM) potential is zero. 
In contrast to the vanishing of gravitational mass, it can be shown that the wormhole solution has finite inertial mass \cite{bib2}, 
thus the equivalence principle is violated.
The vanishing of gravitational mass can be viewed as a result of screening by the scalar field. It is expected that, by
forcing the scalar field to be short-range, the screening effect will be reduced, and the solution will have a finite gravitational mass.
At distances much greater than the bubble radius, the scalar field will be negligible, so the 4-D metric must approach the Schwarzschild 
solution, guaranteeing adherence to the equivalence principle.

The formalism described above can be derived from the least-action principle, with action
\begin{equation}
\label{Eq:action}
S_{KK} = \int d^4 x \sqrt{-g}  \mathcal{L}~,
\end{equation}
where $g$ is the determinant of $g_{\mu \nu}$, and the Lagrangian density is
\begin{equation}
\label{Eq:Lagrangian_density}
 \mathcal{L} =  e ^\Psi (R - \Psi_{; \alpha}^{~\alpha} - \Psi_{, \alpha}\Psi_;^{~ \alpha}) ~,
\end{equation}
where $R$ is the curvature scalar. Setting the variation of $S_{KK}$ with respect to small changes in $g^{\mu \nu}$ to zero yields the Einstein equation with stress-energy given by 
(\ref{Eq:Energy_Momentum}). Variation with respect to small changes in $\Psi$ gives $R = 0$, which is equivalent to (\ref{Eq:psi_wave_eq}).
The Lagrangian density has derivatives up to, and not exceeding, second order, and falls within the class of  Horndeski theories \cite{bib8},
which are free of a troublesome instability \cite{bib9}.

\section{Pinch-Off Conditions}
\label{Pinch-Off}
The radius of the closed fifth coordinate vanishes at $r = r_b$, and this is the boundary
between the interior and exterior regions of the wormhole. In order to obtain a bubble solution rather than a wormhole, one wants a geodesically complete
solution that excludes the interior region.  For this to happen, smoothness criteria must be satisfied by 
the metric at $r = r_b$. Essentially the same situation occurs for Witten's \cite{bib4} growing ``Bubble of Nothing"
and the Gross-Perry \cite{bib2} magnetic monopole. In order to understand the needed conditions in graphic 
terms, consider the ($r,~x^5$) surface defined for fixed values of the time and angular 
coordinates with line element
\begin{equation}
\label{Eq:line_element_2D}
ds^2 = g_{rr}dr^2 + g_{55} a^2 d \phi^2~.
\end{equation}
This two-dimensional curved space can be embedded in a three-dimensional Euclidean
space using cylindrical coordinates $z, \rho,~\phi$ with line element
\begin{equation}
\label{Eq:line_element_3D}
ds^2 = dz^2 + d \rho^2 + \rho^2 d\phi^2~.
\end{equation}
Note that the coordinate $\phi$ is associated with the fifth dimension and is
not one of the angular coordinates implied in the notation $d \Omega^2$ in (\ref{Eq:line_element_5D}).
The desired two-dimensional surface is obtained by constraining the radial coordinate
as $\rho = a\sqrt{g_{55}}$.
Comparing (\ref{Eq:line_element_2D}) and (\ref{Eq:line_element_3D}), one finds the relation between
$\rho$ and $z$ to be
\begin{equation}
\label{Eq:dzdrho}
\frac {dz}{d\rho} = [\frac{4 g_{rr} g_{55}} {a^2 (d g_{55} / dr)^2} - 1]^{1/2}~.
\end{equation}

\begin{figure}[!ht]
\centerline{\includegraphics[width=10.0cm] {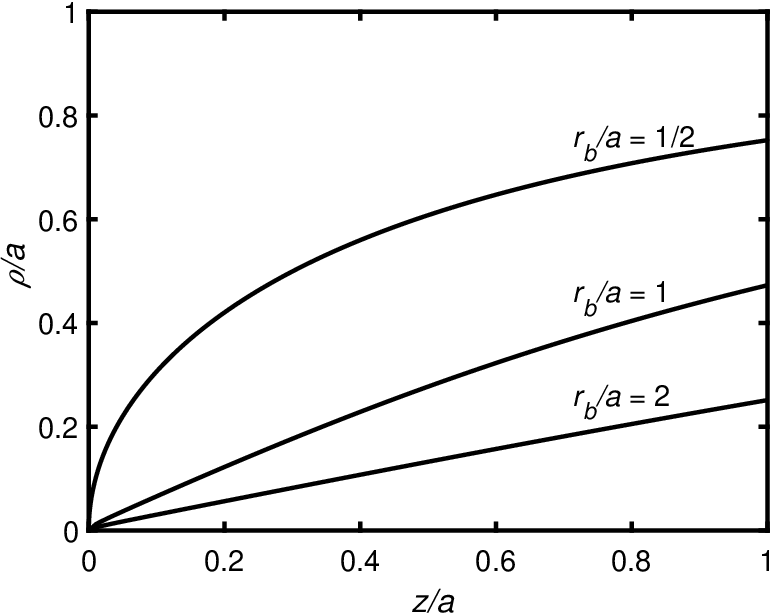}}
\caption{Boundary of axisymmetric ($r,~x^5$) surface embedded in three-dimensional 
($z,~\rho,~\Phi$) Euclidean space. The curves are labeled by the ratio of bubble radius,
$r_b$, to radius, $a$, of the closed fifth dimension.}
\label{fig:1}       
\end{figure}

Figure 1 shows the outline of the embedded, axisymmetric surface. These curves serve to show that the surface has a cusp of 
infinite curvature at $r = r_b$ ($z = 0$) unless 
the pinch-off radius, $r_b$, takes on the value $a/2$. This result is obtained from
(\ref{Eq:dzdrho}) by noting that the condition $dz/d\rho = 0$ must be satisfied at $r = r_b$
if the space is to have finite curvature at the point of closure. Further, $dz^2 /d^2\rho$
must be finite at $r = r_b$. These conditions will ensure that geodesics that originally passed
through $r = r_b$ to the interior region of the wormhole will now turn smoothly at this point 
and return to the exterior region rather than being chopped off. As they touch the former boundary,
$r = r_b$, the coordinate $\phi$ will undergo a step change of $\pi$, but this coordinate
discontinuity does not imply a discontinuity in the geodesic, being analogous to passage through
the North Pole on a line of constant longitude.
In addition to ensuring the vanishing of the first derivative  $dz/d\rho = 0$ at $r = r_b$, the condition
$r_b = a/2$ gives a  finite second derivative $dz^2/d^2\rho = 2/a$. 
It can be shown that, of the family of wormhole solutions in Wesson \cite{bib3}, only the
solution considered here can satisfy the pinch-off criteria.

The bubble solution discussed above violates the equivalence principle and is unstable, with radius initially expanding exponentially with
time \cite{bib7}. The problem with the equivalence principle will be solved by adding a mass term to the equation of motion for the
scalar potential. Stabilization is an issue for the Bubble of Nothing \cite{bib10}
and the Randall-Sundrum model with non-compact fifth dimension \cite{bib11}. Stability questions are not treated here,
but it is hoped that the assumed scalar mass will ultimately be shown to be a necessary result of quantum interactions \cite{bib12},
and that this mass will give bubble stability.

\section{Imposing Mass on The Scalar Field}
\label{Mass_Scalar}
The scalar field will be endowed with mass by adding a term to (\ref{Eq:psi_wave_eq})
\begin{equation}
\label{Eq:psi_wave_eq_mass}
\Psi_{;\mu}^{~~\mu}  + \Psi_{,\mu} \Psi_;^{~\mu} = m^2 \Psi~. 
\end{equation}
This violates five-dimensional covariance and requires a modification of the four-dimensional
stress-energy tensor to restore conservation. The new 4-D stress-energy tensor is assumed to have the form
\begin{equation}
\label{Eq:Energy_Momentum_mass}
T_{\mu\nu}  = \Psi_{;\mu\nu} + \Psi_{,\mu}\Psi_{,\nu} + P g_{\mu \nu}~. 
\end{equation}
As noted earlier, the 4-D metric is denoted $g_{\mu \nu}$. Imposing conservation
\begin{equation}
\label{Eq:Energy_Momentum_cons}
{T_{\mu\nu;}}^\nu  = 0 ~,
\end{equation}
and using the commutation relation
\begin{equation}
\label{Eq:commutation}
{\Psi_{;\nu\mu}}^\nu  -{{\Psi_{;\nu}}^\nu} _\mu  = R_{\mu \nu}{\Psi_;}^\nu ~,
\end{equation}
together with (\ref{Eq:psi_wave_eq_mass}), the following equation is obtained:
\begin{equation}
\label{Eq:DE_P}
P_{,\mu} - P \Psi_{,\mu}  = - m^2 (1 + \Psi / 2) \Psi_{,\mu}~.
\end{equation}
The general solution is
\begin{equation}
\label{Eq:P}
P = \frac{m^2} 2 (\Psi + 3) + b e^\Psi~.
\end{equation}
Choosing $b = -3 m^2/2$ so that the term added to the stress-energy tensor approaches zero as the scalar field approaches zero, the final expression for the conserved stress-energy tensor is
\begin{equation}
\label{Eq:Energy_Momentum_mass_2}
T_{\mu\nu}  = \Psi_{;\mu\nu} + \Psi_{,\mu}\Psi_{,\nu} + \frac{m^2} 2 (\Psi + 3 - 3 e^\Psi ) g_{\mu \nu}~.
\end{equation}
Einstein's equation becomes
\begin{equation}
\label{Eq:Einstein_2}
R_{\mu\nu}  = \Psi_{;\mu\nu} + \Psi_{,\mu}\Psi_{,\nu} + \frac{m^2} 2 (-2 \Psi - 3 + 3 e^\Psi ) g_{\mu \nu}~.
\end{equation}

The author has been unable to find a Lagrangian density that would yield the 4-D expressions given above. This suggests, but does not prove, that the present formalism is not a Horndeski theory \cite{bib8}. Reference \cite{bib9} notes  
that counterexamples show that non-Horndenski theories can be stable.

The metric components in spherical coordinates will be represented by their logarithms according to the usual definitions
\begin{equation}
\label{Eq:Phi}
g_{tt} = -e^{2 \Phi}~,
\end{equation}
and
\begin{equation}
\label{Eq:Lambda}
g_{rr} = e^{2 \Lambda}~.
\end{equation}
After evaluating covariant derivatives, and after some algebraic manipulation, (\ref{Eq:psi_wave_eq_mass}) and (\ref{Eq:Einstein_2}) lead to the equations
\begin{equation}
\label{Eq:Phi_rr}
\Phi_{,rr} = - \frac 1 r \Phi_{,r} (1 + e^{2 \Lambda}) - (r \Phi_{,r} - 1)S~,
\end{equation}
\begin{equation}
\label{Eq:Lambda_r}
\Lambda_{,r} = \frac {{\Phi_{,r}}^2 + (1 + r \Phi_{,r})(1 - e^{2 \Lambda})/r^2 + Q } {2/r + \Phi_{,r}}~,
\end{equation}
and
\begin{equation}
\label{Eq:psi_r}
\Psi_{,r} = \frac {-2 \Phi_{,r}/r +3 m^2 e^{2 \Lambda}(\Psi + 1 -e^\Psi) /2  + (e^{2 \Lambda}-1)/r^2} {2/r + \Phi_{,r}}~,
\end{equation}
with
\begin{equation}
\label{Eq:S}
S =  \frac{m^2} 2 e^{2 \Lambda}(-2\Psi - 3 + 3 e^\Psi)~,
\end{equation}
and
\begin{equation}
\label{Eq:Q}
Q = \frac{m^2} 2 e^{2 \Lambda} \Psi + (1 + r \Phi_{,r})S~.
\end{equation}
Equations (\ref{Eq:Phi_rr}) - (\ref{Eq:psi_r}) comprise a fourth-order system of ordinary differential equations. The fourth equation is the trivial $\Phi_{,r} = D$, with $D$ replacing $\Phi_{,r}$ in  (\ref{Eq:Phi_rr}), (\ref{Eq:Lambda_r}), (\ref{Eq:psi_r}), (\ref{Eq:Q}), and with $\Phi_{,rr}$ in (\ref{Eq:Phi_rr}) becoming $D_{,r}$. The unknown variables are $D$, $\Phi$, $\Lambda$, and $\Psi$.
A bubble solution must meet two conditions: The initial dependence of $\Lambda$ and $\Psi$ for $r$ near $r_b$ must satisfy the pinch-off conditions, and the asymptotic behavior of $\Phi$ and $\Lambda$ for $r >> r_b$ must be that of a massive body, that is, must approach the Schwarzschild solution. As 
the above equations governing radial dependence are too complicated for analytical solution and straightforward interpretation, these conditions will be examined by use of a numerical solution.

\section{Computation}
The numerical solution of (\ref{Eq:Phi_rr}) - (\ref{Eq:psi_r}) employs the shooting method, with initial conditions varied until the scalar field approaches zero asymptotically. In order to guarantee that the pinch-off conditions are met, both $\Lambda$ and $\Psi$ are required to be equivalent to (\ref{Eq:hrr}) and (\ref{Eq:h55}) as $r/r_b \rightarrow 1$. That is,
\begin{equation}
\label{Eq:Lambda_initial}
\Lambda \rightarrow -\frac 1 2 \ln (1-r_b/r) ~,
\end{equation}
and
\begin{equation}
\label{Eq:psi_initial}
\Psi \rightarrow \frac 1 2 \ln (1-r_b/r) ~.
\end{equation}
As these limits are infinite, the numerical computation is started for $r$ slightly larger than the bubble radius $r_b$.
With the idea of screening of the gravitational field by the scalar field in mind, it is assumed that $-\Phi$ for $r$ near $r_b$ has a value much smaller than $\Lambda$. Consequently, shooting consists of trying various large negative values of $\Phi_{,r}$ at the initial  $r/r_b$ until one finds a solution for which the scalar field asymptotically approaches zero.  Inspection of (\ref{Eq:Phi_rr}) - (\ref{Eq:psi_r}) 
shows that any solution can be altered by adding a constant to $\Phi$. This is simply a scaling of the time variable, and this freedom is used to force 
$\Phi \rightarrow -\Lambda$ as $r/r_b \rightarrow \infty$ as in the Schwarzschild solution. 

\begin{figure}[!ht]
\centerline{\includegraphics[width=10.0cm] {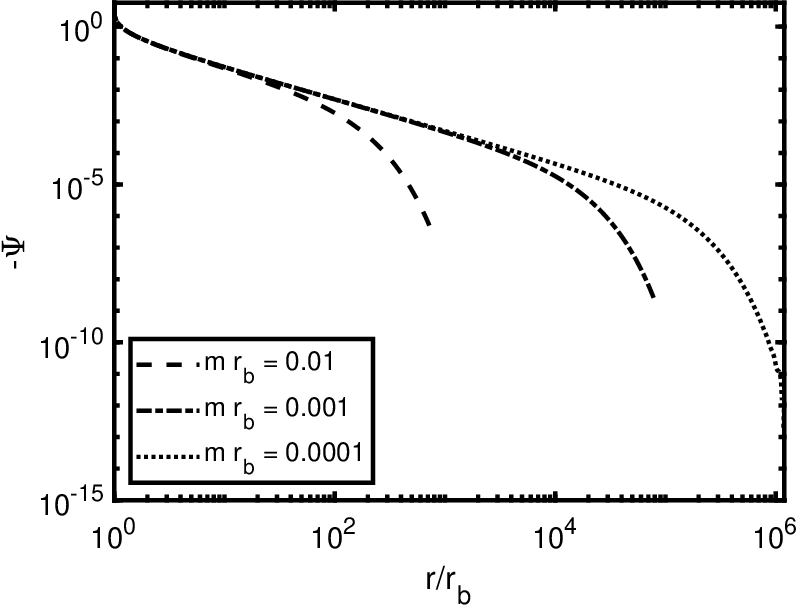}}
\caption{Radial falloff of the scalar field for three choices of the scalar mass $m$. (The scalar field is negative everywhere.)}
\label{fig:2}       
\end{figure}

Figure \ref{fig:2} shows the radial dependence of the scalar field $\Psi$ with falloff from long-range behavior setting in at shorter ranges as scalar mass increases.

\begin{figure}[!ht]
\centerline{\includegraphics[width=10.0cm] {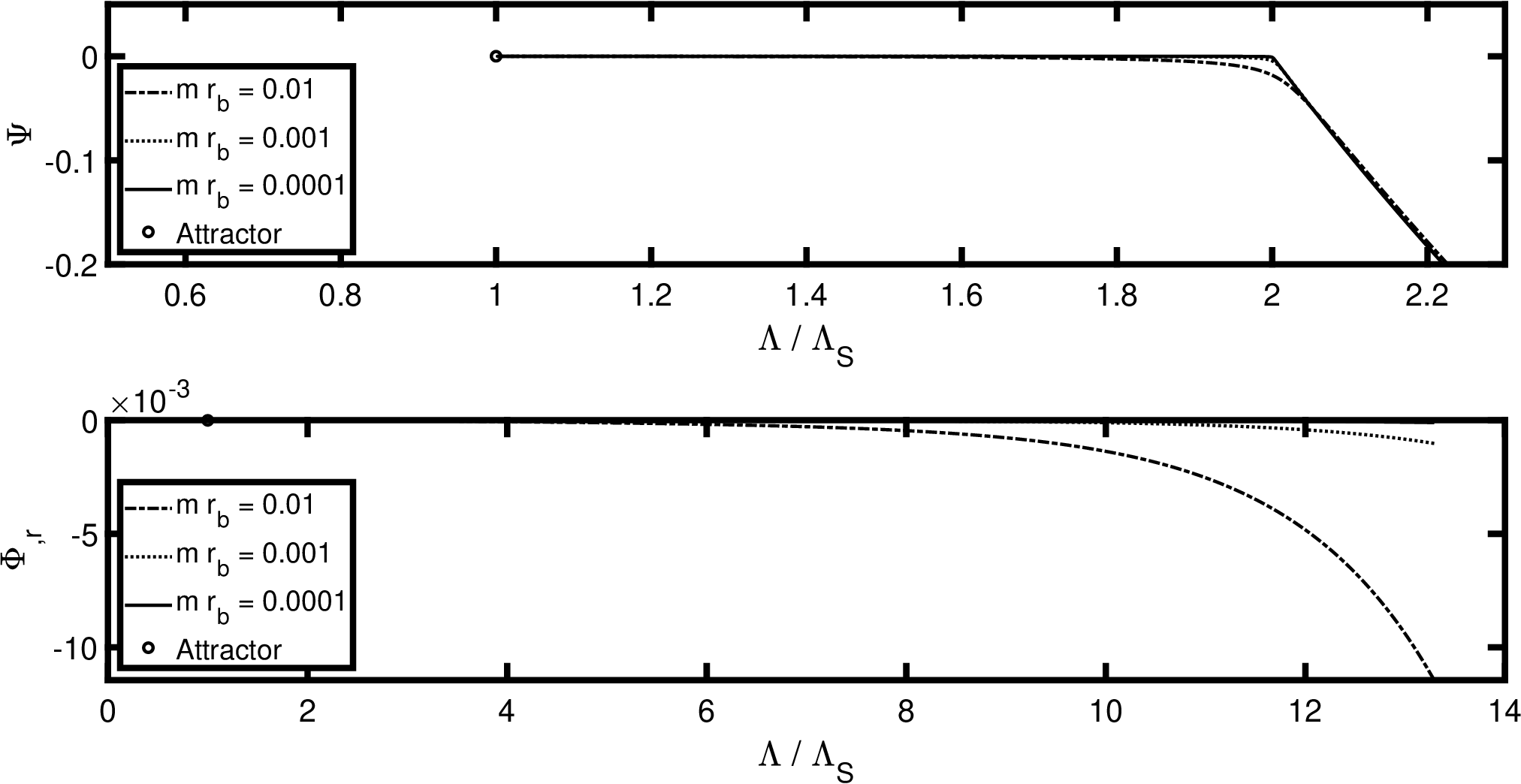}}
\caption{Approach of the bubble solution for three choices of scalar mass toward an attractor corresponding to a Schwarzshild metric with radius $r_b /2$.}
\label{fig:3}       
\end{figure}

The author expected the bubble mass would be a function of the scalar field mass, but computations strongly suggest that the bubble mass is independent of
the scalar field mass. This indicates that the soluion of the system of equations  (\ref{Eq:Phi_rr}) - (\ref{Eq:psi_r}) approaches an attractor as $r$ increases. 
The attractor is the long-range Schwazshild solution. In support of this conclusion Fig. \ref{fig:3} shows this behavior in plots of
$\Psi$ and $\Phi_{,r}$ vs. $\Lambda / \Lambda_S$, where $\Lambda_S = -\ln[1-r_b/(2 r)]/2$ is from the Schwarzshild solution with a Schwarzshild radius of $r_b /2$.

In both Figs. \ref{fig:2} and \ref{fig:3} the scalar mass is specified in terms of the dimensionless parameter $m r_b$. This parameter can be converted to 
standard mass units through the expression  $m_{\rm{scalar}} = m r_b \sqrt{\alpha} m_P$, where $m_P$ is the Planck mass.

The result that the 
Schwarzschild radius is smaller than the bubble radius is of no concern, as the bubble is a complicated object with extended scalar field and no event horizon.
Appendix \ref{Attractor} gives further justification for the interpretation of this result in terms of an attractor.

\section{Conclusions}
The bubble radius $r_b$ is one-half the K-K radius, $a$, so the bubble mass follows from the well-known expression for the K-K radius \cite{bib13}:
 \begin{equation}
\label{Eq:K-K_radius}
a = \frac 2 {\sqrt{\alpha}} {\it l}_P ~,
\end{equation}
where $ {\it l}_P$ is the Planck length, and  $\alpha$ is the fine-structure constant.
The Schwarzschild radius of the bubble is one-half the bubble radius, $r_b$, independent of the choice of $m$, at least over the range
$0.001 < m r_b < 0.00001$. A computation for  $m r_b =  0.01$ suffers from numerical problems, but $m r_b = 0.00001$ presents
no difficulty. No curves for this case are plotted in Fig. \ref{fig:3} because they are difficult to distinguish from the $m r_b = 0.001$ curve 
in the upper panel. The choice $m r_b = 0.01$ places the scalar-field mass in the vicintity of the Grand-Unification scale, while the value $m r_b = 0.00001$ is in the range of some proposed inflaton masses  \cite{bib14,bib15}.
Using the relation between the Schwarzschild radius and gravitational mass, the bubble mass is
 \begin{equation}
\label{Eq:soliton_mass}
m_{\rm{bubble}} =  \frac {m_P} {4 \sqrt{\alpha}} .
\end{equation}
The bubble is exteremly heavy, about three times the Planck mass. The bubble would be very difficult to detect, as its only interaction is via gravity.
If questions regarding stabiity were resolved, the bubble might be considered a dark-matter candidate.

\appendix

\section{Attractor}
\label{Attractor}
Equations (\ref{Eq:Phi_rr}) - (\ref{Eq:psi_r}) show that there are four variables of interest, $\Psi$, $\Phi_{,r}$, $\Phi$, and $\Lambda$.
An attractor is a sub-region of the variable space which the evolving solution approaches, independent of initial conditions, provided these initial conditions
fall within a subspace termed the ``basin of attraction". An objection to the use of the term ``attractor" in the present work might be raised, because the system (\ref{Eq:Phi_rr}) - (\ref{Eq:psi_r}) contains the scalar mass, $m$, as a parameter, and this parameter is also varied along with the initial conditions. 
This concern can be eliminated by using a new radial coordinate $\hat{r} = m r$ for which the system of ordinary differential equations is independent
of $m$, which then appears in the initial conditions, as the variables that are derivatives with respect to the radius now include a factor $1/m$.  

It is more convenient, however, to revert to the original variables when descibing the attractor and basin of attraction. The attractor is formed from the 
Schwarzschild solution for $r >> r_b$, with Schwarzschild radius $r_b/2$. Then the attractor is defined by $\Lambda = r_b/(4 r)$, 
$\Phi = -\Lambda + \rm{constant}$,
 $\Phi_{,r} = -r_b/(4 r^2)$, and $\Psi = 0$. As noted earlier, $\Phi$ may have an additive arbitrary constant, as only $\Phi_{,r}$ appears in 
 (\ref{Eq:Phi_rr}) - (\ref{Eq:psi_r}). In Fig. \ref{fig:3}, the attractor is represented by a single point in a space comprised of variables
$\Psi$, $\Lambda/\Lambda_S$, and $\Phi_{,r}$, where $\Lambda_S$ is from the Schwarzschild solution with radius $r_b/2$. 
In assessing the curves in Fig. \ref{fig:3} it is important to note that the abscissas have a counterintuitive direction,
with large  $\Lambda/\Lambda_S$ corresponding to regions near the bubble and with the attractor point corresponding to the limit $r \rightarrow \infty$.
The variables 
 $\Phi = -\Lambda + \rm{constant}$ and $\Phi_{,r} = -r_b/(4 r^2)$ need not be considered in describing the attractor, as it is easily shown that the 
system of differential equations guarantees these conditions  will be satisfied if the scalar field tends toward zero at large $r$, as shown in Fig. \ref{fig:2}.

The basin of attraction can be defined as a two-dimensional surface in  $\Psi$, $\Lambda/\Lambda_S$, $\Phi_{,r}$ space. 
The upper panel of Fig. \ref{fig:3} is a top view of
slices through the basin, with the slices depending on $\Lambda/\Lambda_S$ as shown in the lower panel.
Together the two panels show three trajectories moving through
the basin to the attractor. Each of these trajectories corresponds to a particular scalar field mass, as noted in the figures' legends. No attempt has been made to
determine the limits of the basin with respect to $m$, as this would require substantial numerical effort. In this connection, it 
is likely that failure of the curves in the upper panel to overlap more perfectly for large $\Lambda/\Lambda_S$ may be due to 
numerical error. Better overlap is expected, as the numerical solution should approach the massless bubble solution as $r \rightarrow r_b$. 
The sharp turn of the trajectories near  $\Lambda/\Lambda_S = 2$ in the upper panel of Fig. \ref{fig:3} shows that this is the region where the solution 
transitions from that of the original bubble to that of a gravitating body.


%
%

\end{document}